\documentstyle[12pt,aaspp4]{article}

\begin{document}

\title{Submillimeter Continuum Emission in the $\rho$ Ophiuchus Molecular 
Cloud: Filaments, Arcs, and an Unidentified Far-Infrared Object} 
\author{Christine D. Wilson\altaffilmark{1,2},
Lorne W. Avery\altaffilmark{3,4},
Michel Fich\altaffilmark{5},
Doug Johnstone\altaffilmark{6},
Gilles Joncas\altaffilmark{7},
Lewis B. G. Knee\altaffilmark{8},
Henry E. Matthews\altaffilmark{3,4},
George F. Mitchell\altaffilmark{9},
Gerald H. Moriarty-Schieven\altaffilmark{3,4},
Ralph E. Pudritz\altaffilmark{1}}
\bigskip

\altaffiltext{1}{Department of Physics and Astronomy, McMaster University,
Hamilton, Ontario L8S 4M1 Canada} 
\altaffiltext{2}{Division of Physics, Mathematics, and Astronomy,
Caltech 105-24, Pasadena CA 91125 U.S.A.}
\altaffiltext{3}{Herzberg Institute of Astrophysics, 5071 W. Saanich Road, 
Victoria B.C. V8X 3M6 Canada}
\altaffiltext{4}{Joint Astronomy Centre, 660 N. A'ohoku Place,
Hilo HI 96720 U.S.A.}
\altaffiltext{5}{Department of Physics, University of Waterloo,
Waterloo ON N2L 3G1 Canada}
\altaffiltext{6}{Canadian Institute for Theoretical Astrophysics, 60 St. George
St., Toronto ON M5S 3H8 Canada}
\altaffiltext{7}{D\'epartement de Physique, Universit\'e Laval, Sainte-Foy
PQ G1K 7P4 Canada}
\altaffiltext{8}{National Research Council Canada, Herzberg Institute
of Astrophysics, Dominion Radio Astrophysical Observatory, P.O. Box 248,
 Penticton B.C. V2A 6K3 Canada}
\altaffiltext{9}{Department of Astronomy and Physics, Saint Mary's University,
Halifax NS B3H 3C3 Canada}

\begin{abstract}

New wide-field 
images of the $\rho$ Ophiuchus molecular cloud at 850 and 450 $\mu$m
obtained with SCUBA on the James Clerk Maxwell Telescope reveal a wide
variety of large-scale features that were previously unknown. Two
linear features each 4$^\prime$ (0.2 pc) in length extend to the north
of the bright emission region containing SM1 and VLA 1623. These features
may correspond to  the walls of a previously unidentified outflow cavity,
or the boundary of a photon dominated region powered by a nearby B star.
A previously unidentified source is located in the north-east region
of the image. The properties of this source (diameter $\sim 5000$ AU, mass
$\sim 0.3-1$ M$_\odot$) suggest that it is a pre-protostellar core.
Two arcs of emission are seen in the direction of
the north-west extension of the VLA 1623
outflow. The outer arc appears relatively smooth
at 850 $\mu$m and is estimated to have a mass of $\sim 0.3$ M$_\odot$, while
the inner arc breaks up into a number of individual clumps, some of which
are previously identified protostars. 

\end{abstract}


\section{Introduction}

Young star-forming regions can encompass a wide variety of sources and
phenomena, from the pre-protostellar clumps and the quiescent
filamentary structure of the parent molecular cloud, 
to low-mass protostars with
their associated disks, jets, and outflows, to young massive stars powering
extensive photon-dominated regions. In the nearest clouds, where we
can achieve the best spatial resolution for studying small-scale phenomena
such as circumstellar disks and for disentangling crowded
clusters of protostars, the largest outflows may extend over
tens of arcminutes (i.e. \markcite{d95}Dent, Matthews, \& Walther 1995),
while the molecular cloud itself may approach a degree or more in size
(i.e. \markcite{m86}Maddalena et al. 1986). 
Thus, high resolution imaging over large fields of view
is necessary for a complete understanding of such regions.

Many of these phenomena are best traced by continuum emission, particularly
the high-density cores and envelopes associated with the early
stages of star formation. Given the limited sensitivity of single-pixel
bolometers, most surveys have targeted sources previously identified
from larger area surveys in the near- or far-infrared 
(i.e. \markcite{a94}Andr\'e
\& Montmerle 1994). Although such observations can give us a good
understanding of the properties of young protostars, they cannot
reveal the large-scale structures associated with the protostars
and, in addition, may miss the coldest and most crowded objects.
For example, a large-area survey of the $\rho$ Ophiuchus molecular
cloud using the IRAM bolometer array (\markcite{m98}Motte, Andr\'e,
\& Neri 1998) was able to identify a number
of weaker sources located in and around the bright complex of
emission associated with SM1 and VLA 1623 (\markcite{a93}Andr\'e,
Ward-Thompson, \& Barsony 1993). With the advent of SCUBA 
(\markcite{h98}Holland et al. 1999) on the James Clerk Maxwell Telescope,
we now have the opportunity to survey large areas of the sky
in submillimeter continuum emission routinely. Sensitive, large-area
surveys that extend beyond the known regions of
bright emission are likely to reveal previously unknown features, as was
clearly demonstrated by recent maps 
of the Orion A molecular cloud (\markcite{j98}Johnstone \& Bally 1999).
In this Letter, we present the first results from a large-area
unbiased survey of the $\rho$ Ophiuchus molecular cloud at 850 and
450 $\mu$m.

\section{Observations and Data Reduction}

We observed a $20 \times 20^\prime$ region of the $\rho$ Ophiuchus molecular
cloud roughly centered on the Ophiuchus A core on 1998 July 10 and 11
with the bolometer array SCUBA (\markcite{h98}Holland
et al. 1999) at the James Clerk Maxwell Telescope. We observed in 
standard scan-mapping
mode simultaneously at 850 and 450 $\mu$m. The region was broken down
into four fields each measuring roughly
$10^\prime \times 10^\prime$. In this paper, we present data for only the 
north-east field centered on 16:26:32 -24:22:30 (J2000); 
all subsequent discussion refers to this single field.

The field was mapped six times each night, 
with chop throws of 20$^{\prime\prime}$,
30$^{\prime\prime}$, and 65$^{\prime\prime}$ oriented in either the right
ascension or declination directions. Each chop throw and direction
was repeated
at a different time on the two nights so that the scan direction would
have a different orientation on each night. The data were acquired with 
3$^{\prime\prime}$ sampling.
The atmospheric
optical depth, measured by sky dips approximately once an hour,
 was quite constant on the first night, with the value at zenith
being 0.10 at 850 $\mu$m and 0.43  at 450 $\mu$m
during the time these observations were obtained. On the second night,
the optical depth improved steadily over the course of the 
observations, decreasing
from 0.35 to 0.24 at 850 $\mu$m and from 1.77 to 1.33  at 450 $\mu$m. 
The pointing was checked every 45 minutes using the bright
source IRAS 16293-2422 and was stable to better than 2$^{\prime\prime}$.
The beam sizes measured from observations of Uranus obtained with
the same observing method were 15.1$^{\prime\prime}$
(FWHM) at 850 $\mu$m and 9.5$^{\prime\prime}$ at 450 $\mu$m. The presence of
an extended error beam can be seen at both wavelengths, with a FWHM
of 36$^{\prime\prime}$ at 850 $\mu$m and 72$^{\prime\prime}$ at 450 $\mu$m.
At 850 $\mu$m the peak of the error beam is 5.6\% of that of the main beam,
while at 450 $\mu$m the peak of the error beam is 3.6\%.
The observations
of Uranus were also used to determine the calibration factor, which
was 198 Jy Volt$^{-1}$ at 850 $\mu$m and 629 Jy Volt$^{-1}$ at 450 $\mu$m.

The individual scan maps were processed using the standard SCUBA software
(\markcite{h98}Holland et al. 1999) 
into six independent dual-beam maps at each wavelength, with pixel size
3$^{\prime\prime}$. The maps have not been corrected for sky noise.
The six dual-beam maps were exported to the MIRIAD software package 
(\markcite{s95}Sault, Teuben, \& Wright 1995)
to be analyzed using the maximum entropy algorithm. Each map was shifted
by one half the chop throw and the six shifted maps were averaged to
produce a ``dirty map'' of the region showing the pattern of the 6 chop
throws. We then created an ideal ``beam'' consisting of the sum of
one positive gaussian and six negative gaussians, each with a FWHM equal
to that of the observed main beam, with the six negative gaussians offset
from the origin by the six chop throws, and with peak amplitude one-sixth
that of the positive gaussian. The contribution of the error beam
was ignored in this analysis; the effect of this will be to
increase artificially the integrated fluxes of extended sources
above what would be observed with an ideal beam.
For a gaussian source
with a full-width half-maximum diameter of 35$^{\prime\prime}$, the
net effect would be to increase the total 450 $\mu$m 
flux inside a 40$^{\prime\prime}$
radius annulus by about 30\%. However, peak fluxes will be unaffected
by this problem. The dirty map and the beam were then
used as inputs to the miriad routine ``maxen'' to create a maximum
entropy restored image of the scan-mapped field. Clean boxes were used
to isolate the large negative bowls around the bright sources SM1, SM1N, SM2,
and VLA 1623, which improved the image restoration. The final images
still contain residuals of the negative chop throws at a level of 4-6\% 
of the peak flux in the map.
However,  the negative bowls are much reduced in area and depth 
compared to the images obtained with the ``Emerson II'' technique
(\markcite{h98}Holland et al. 1999) and, thus, are
more useful for identifying large scale features in the map.
The typical rms noise
in the final maps far from the negative bowl 
is estimated to be 30 mJy beam$^{-1}$ at 850 $\mu$m and
250 mJy beam$^{-1}$ at 450 $\mu$m.

\section{A Wealth of Previously Unknown Structures}

\subsection{Linear Features}

Color images of the 850 $\mu$m and 450 $\mu$m maps of the region
around the Ophiuchus A
cloud core are shown in Figure~\ref{fig-1}. One of the 
most striking aspects  of these images is the two linear emission
features extending from the northern tip of the bright complex of
emission associated
with the protostellar cores SM1, SM1N, SM2, and VLA 1623 
(\markcite{a93}Andr\'e et 
al. 1993). The two linear features appear to intersect near 16:26:29 -24:22:45 
(J2000) and extend for at least 4$^\prime$ (0.2 pc) to the north-east
and the north-west. The peak surface brightness in these linear features
is 0.6-0.7 Jy beam$^{-1}$ at 850 $\mu$m near the SM1 complex,
while at large distances from the complex the peak surface brightness
falls to $\sim 0.1-0.2$ Jy beam$^{-1}$. These linear features are
so weak that they can only be identified with the high sensitivity of SCUBA
and in wide-field images which allow their large linear extent to be traced.
Similar linear features have been seen recently in Orion A
(\markcite{j98}Johnstone \& Bally 1999).

In the vicinity of the intersection of the two features, there is strong CO
emission which is blueshifted by 1-2 km s$^{-1}$ compared to the
central velocity of the emission around VLA 1623 
(D. Koerner, private communication). 
Thus, it is possible that these two linear features
trace the walls of a previously unidentified outflow cavity.
For comparison, the walls of the south-east outflow lobe created by VLA 1623
are faintly visible in Figure~\ref{fig-1}.
This outflow would have  an  opening angle of 
56$^o$, comparable to that observed for the L1551
outflow (\markcite{ms87}Moriarty-Schieven et al. 1987). 
At the apex of the ``V'' formed by the two features there is a weak compact
source, seen most easily in the 20\% contour of Figure~\ref{fig-1}b.
This source has been  identified previously at 1.3 mm as A-MM6 
(\markcite{m98}Motte et al. 1998) and
could be the protostellar driver of the 
outflow (Figure~\ref{fig-1}). 
However, there
are many other possible interpretations for these linear features.
For example, they could trace two independent outflows, which would then
have collimation factors of at least 7-10. Two independent
outflows emanating from a single position have
been observed in several other protostellar
sources (i.e. L723, \markcite{a91}Anglada  et al. 1991;
IRAS 16293-2422, \markcite{m92}Mundy et al. 1992). 
Another possibility is that these structures are being externally heated
by the photon-dominated region (PDR) produced by the young B3 star S1, which
lies approximately 1.5$^\prime$ east of SM1N (i.e., see \markcite{m98}Motte
et al. 1998). In this scenario, the north-east
feature could mark the cavity wall containing gas and dust
swept up by the PDR; indeed, the location of this feature matches well
the edge of the PDR seen in ISO images (\markcite{a96}Abergel et al. 1996).
The north-west feature might be material
associated
with a second outflow recently identified in this region
(Kamazaki et al. 1998), which could be heated from the outside
by the PDR.
The presence of a weak linear
feature along the southern edge of the map, which appears to lie along the
northern edge of CO outflow associated with VLA 1623 (\markcite{a90}Andr\'e 
et al. 1990), provides some support for this interpretation. 
Clearly, sensitive CO observations in the region of
these new linear filaments, as well as continuum observations
of regions which are not illuminated by a nearby massive star, would help 
to distinguish between these different interpretations.

Assuming a dust temperature of 30 K, 
a distance to $\rho$ Ophiuchus of 160 pc, and a dust opacity coefficient
$\kappa = 0.02$ cm$^{2}$ g$^{-1}$ at 850 $\mu$m (i.e.
\markcite{m98}Motte et al. 1998), we can estimate the total
mass (gas plus dust) contained in these linear features.
The 850 $\mu$m 
fluxes in the peak regions near the SM1 complex correspond to
masses within a 15$^{\prime\prime}$ beam of 0.04-0.05 M$_\odot$.
However, the presence of extended low level emission associated with
the SM1 complex means that these masses are likely upper limits to the
true masses. The masses in the strongest emission regions 2-4$^\prime$
along the linear features 
from this bright complex are each of order 0.01 M$_\odot$.
The average surface brightness of 0.04 Jy beam$^{-1}$ along the outer
150$^{\prime\prime}$ of the north-east linear feature (excluding the bright
base) translates into a mass of 0.03 M$_\odot$. Thus, the total mass in
the north-east linear feature is likely to be of order 0.1 M$_\odot$.
For comparison, this is similar to 
the mass of 0.09 M$_\odot$ estimated from CO observations of
the IRAS 03282 outflow (\markcite{b91}Bachiller, Martin-Pintado, \& Planesas
1991).
However, the masses of $\sim 0.01$ M$_\odot$
in the individual clumps are substantially
larger than the masses of $10^{-4}$ M$_\odot$ obtained for 
the CO bullets seen in extremely
high velocity outflows (i.e. \markcite{ba90}Bachiller et al. 1990). 

\subsection{A New Compact Source in Ophiuchus}

Another striking feature of the images presented in Figure~\ref{fig-1}
is the presence of four bright compact sources that lie
well outside the bright emission associated with SM1 and VLA 1623.
Three of these sources are the previously identified protostars
EL24, EL27, and GSS26 (i.e. \markcite{a94}Andr\'e \& Montmerle 1994), while the
fourth, located in the north-east corner of our field, appears to be
previously unidentified. We will refer to this north-east
source as SMM16267-2417. (Note that 
\markcite{l90}Loren, Wootten, and Wilking (1990)
detected DCO$^+$ emission $\sim 1^\prime$ north of this source, but the
line emission was too weak for it to be designated as a DCO$^+$ core.)
This source is located at 
16:26:43.5  -24:17:26 (J2000) and has a peak flux of 0.4 Jy beam$^{-1}$
at 850 $\mu$m and 1.9 Jy beam$^{-1}$ at 450 $\mu$m.
Its full-width
half-maximum diameter at 850 $\mu$m deconvolved from the 15.1$^{\prime\prime}$
beam is $26 \times 34^{\prime\prime}$ or $\sim 5000$ AU.
Assuming an uncertainty of 10\% for each point in the observed
450 $\mu$m radial profile, a ``by-eye'' fit between radii of 
9$^{\prime\prime}$ and 24$^{\prime\prime}$ suggests a slope of
$-0.8\pm 0.2$. This slope is comparable to that seen in the
outer portions of the pre-protostellar cores by
\markcite{wt94}Ward-Thompson et al. (1994).
A more complete modeling of the radial profile of SMM16267-2417 will
be presented in a future paper.
No point-like source, or indeed any obvious emission,
can be seen at any waveband in either the IRAS FRESCO or
HIRES images, except at 100 $\mu$m where SMM16267-2417 is
located on the northern slope of another structure to the south.  
Assuming a dust temperature of 10-20 K for SMM16267-2417 and the other
parameters as described above for the linear features,
the 850 $\mu$m flux implies a total mass of 0.3-1 M$_\odot$.
These properties are all 
consistent with SMM16267-2417 being a pre-protostellar
core (i.e. \markcite{wt94}Ward-Thompson et al. 1994). 
However, if
the linear features identified in this region indeed correspond to a
molecular outflow, it is possible that SMM16267-2417 corresponds
to shocked gas within the outflow cavity. 
Here again,
CO and other line observations would help distinguish between these 
various possibilities.

The spectral index $\gamma$, defined such that the flux $S_{\nu}$ is
proportional to  $\nu^{\gamma}$ between 450 and 850 $\mu$m,
is related to the dust index $\beta$ and the dust temperature $T$.
At temperatures below $30\,$K (Ward-Thompson et al. 1994), the
Rayleigh-Jeans assumption is inappropriate, and 
$\gamma$, $\beta$, and $T$ are related by
$$ {S_{450} \over S_{850}} = (850/450)^\gamma 
= {{e^{16.9/T}-1} \over {e^{32.0/T}-1}} (850/450)^{3+\beta}$$
where $S_{450}$ and $S_{850}$ are the fluxes at 450 and 850 $\mu$m.
We have assumed that the gas is optically
thin at 450 $\mu$m, which is appropriate if the flux
originates from material that fills the beam.
Using the integrated fluxes of
2.3 Jy at 850 $\mu$m and 19.3 Jy at 450 $\mu$m (correcting
for the error beam as discussed in \S 2), 
SMM16267-2417 has an average spectral index of 3.3.
The value of $\beta$ derived from these wavelengths is highly dependent
on the assumed dust temperature. 
Assuming 
an uncertainty in the flux ratio of 20\% and a dust
temperature of 20 K, this spectral
index corresponds
to a value for $\beta$ of $2.1^{+0.2}_{-0.4}$. This
value of $\beta$ is consistent with that found on large scales
in the interstellar medium (\markcite{h83}Hildebrand 1983), 
but somewhat larger than
the values of $\beta \sim 1$ that are typically found in compact
cores and circumstellar disks (\markcite{bs91}Beckwith \& Sargent 1991).

\subsection{Arcs and Other Features}

Two curved arcs of continuum emission are clearly visible in
the 850 $\mu$m image of Figure~\ref{fig-1}a to the north-west of the SM1 
complex. The more distant arc (Arc \#2, at 16:26:10, -24:20) has a peak
surface brightness of 0.29 Jy beam$^{-1}$ and an integrated flux of
3.7 Jy. Arc \#2 appears quite smooth and does not obviously
break up into point sources. 
Assuming a
dust temperature of 30 K, the total mass in this arc is 0.3 M$_\odot$. 
This arc may be related to the VLA 1623 outflow; although the outflow
has not been mapped out this far (\markcite{a90}Andr\'e et al. 1990,
\markcite{d95}Dent et al. 1995), if the outflow continues on in the
same direction in the outer portions, it would pass just to the south
of this arc.

The inner arc (Arc \#1, at 16:26:20, -24:23), 
although similar in structure to Arc \#2,
 clearly breaks up into five clumps in
the 850 $\mu$m map. From north to south, these clumps are A-MM4,
LFAM1, A-MM1, A-MM3, and LFAM3
(\markcite{a94}Andr\'e \& Montmerle 1994,
\markcite{m98}Motte et al. 1998). 
(We do not see a clear peak corresponding to A-MM2 in
our 850 $\mu$m images.) At 850 $\mu$m, the emission
from LFAM1 dominates the emission from GSS30-IRS1 and IRS2, although
weak peaks at the approximate location of these sources can be
seen in the higher resolution 450 $\mu$m map. \markcite{m98}Motte et al. (1998)
suggested that the clumps A-MM1 to A-MM3 may be related to the VLA
1623 outflow, which passes through this region; however,
A-MM4 lies well outside the outflow. The presence of Arc \#2
more distant from VLA 1623 but still coincident with the
outflow suggests that in both cases we may be seeing continuum emission
associated with bow shocks in the VLA 1623 outflow.

\acknowledgments

The research of MF, DJ, GJ, GFM, REP, and CDW
is supported through grants from the Natural
Sciences and Engineering Research Council of Canada. The JCMT is operated by
the Joint Astronomy Centre on behalf of the Particle Physics and Astronomy
Research Council of the United Kingdom, the Netherlands Organization
for Scientific Research, and the National Research Council of Canada.

\clearpage

\figcaption[fig1.ps]{(a) 
The 850 $\mu$m image of the Ophiuchus A region
obtained with SCUBA on the JCMT. The color table has been scaled
to emphasize weak features and the resolution of the map is 
15.1$^{\prime\prime}$. The linear features discussed in Section 3.1
originate at approximately
16:26:29 -24:22:45  and extend out to 16:26:40 -24:19:30 for the 
north-east feature, and 16:26:20 -24:18:30 for the north-west feature.
The negative bowl
of emission to the east of the SM1 complex is an artifact of the dual beam
mapping that was not completely removed by the maximum entropy reconstruction.
The contour levels are 15,20,25,...95\% of the peak flux (5.24 Jy beam$^{-1}$).
The solid lines indicate the direction of the VLA 1623 outflow and the 
five sources along Arc \#1 are marked by plus signs.
(b) The 450 $\mu$m image of the Ophiuchus A region
obtained with SCUBA on the JCMT. The color table has been scaled
to emphasize weak features and the resolution of the map is 
9.5$^{\prime\prime}$. 
The contour levels are 15,20,25,...95\% of the peak flux (22.6 Jy beam$^{-1}$).
\label{fig-1}}

\clearpage
 
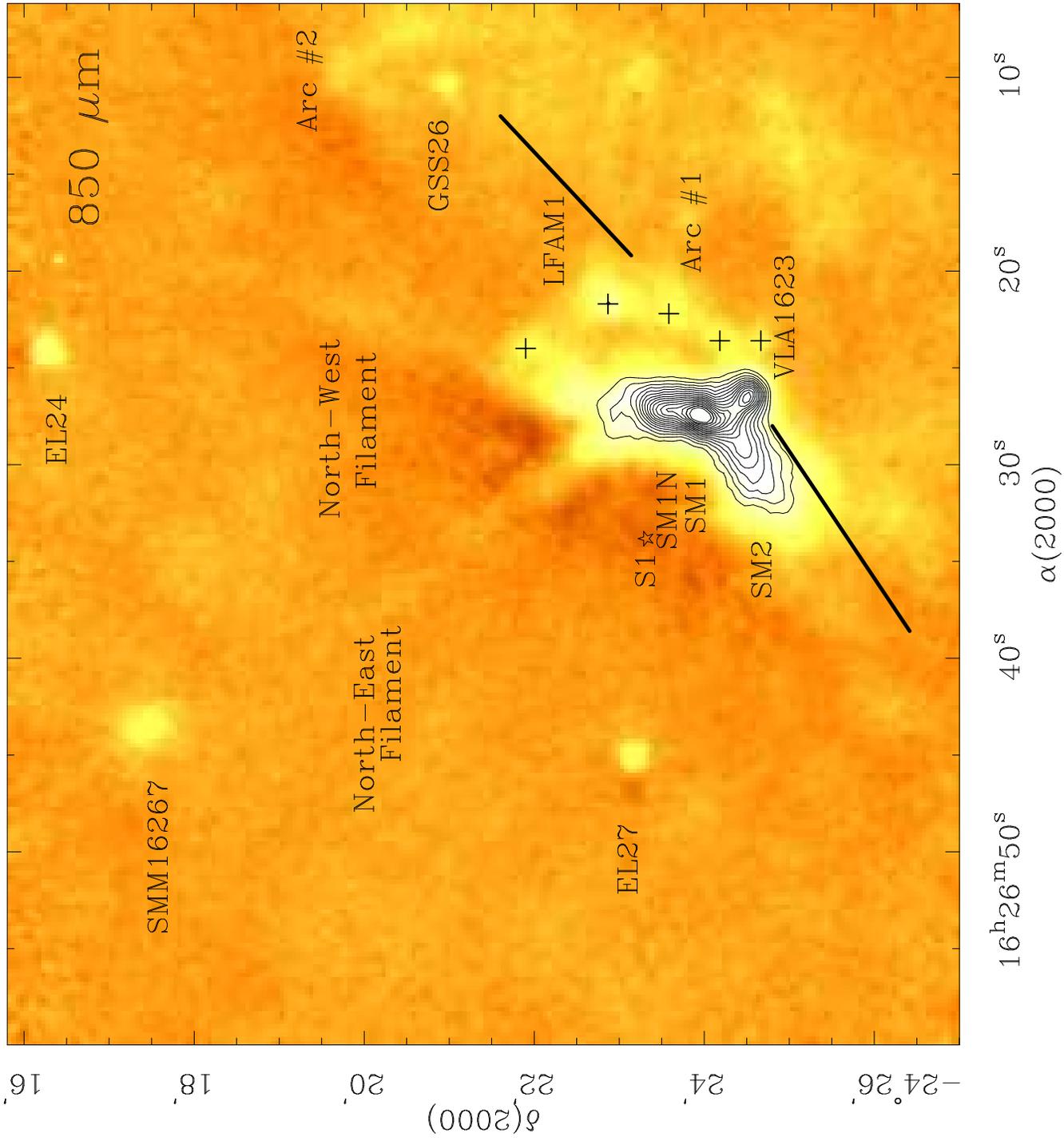
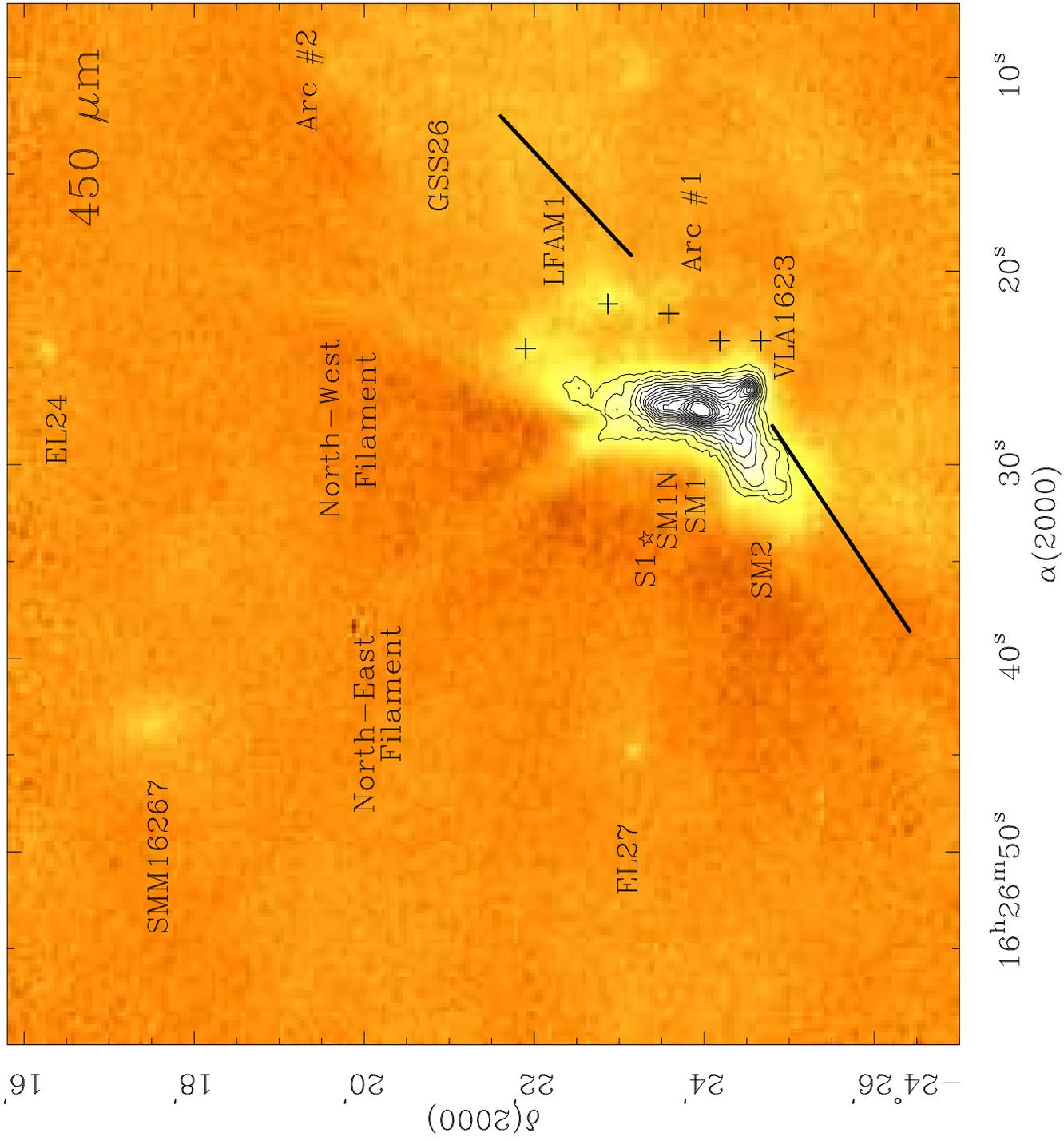
\end{document}